# The unusually strong coronal emission lines of SDSS J1055+5637

**Hartmut Winkler**

Department of Physics, University of Johannesburg, South Africa

E-mail: hwinkler@uj.ac.za

**Abstract**. Many Seyfert galaxies display weak 'coronal' emission features corresponding to [Fe VII], [Fe XI] and [Fe XIV] in their optical spectra, whereas elsewhere these lines seem to be entirely absent. These lines appear to highlight zones in the nucleus irradiated by high-energy photons. The presence of these zones and the conditions therein as determined by the relative line strengths and profiles impose important constraints on the physical models of active galactic nuclei, and Seyferts in particular. In 2009 the discovery was announced of the highly unusual spectrum of SDSS J0952+2143, where the coronal lines are exceptionally strong. This paper presents a second object with abnormally strong coronal features, SDSS J1055+5637. The spectrum, line ratios and related parameters are compared to those of SDSS J0952+2143, three AGN with moderate coronal lines and one where the coronal lines are missing altogether. Possible mechanisms are discussed that may account for the stronger than usual coronal features.

## 1. Introduction

Emission lines associated with highly ionised iron have been noticed in the spectra of some Seyfert galaxies since some of the early investigations of this class of active galactic nuclei (AGN). These have been dubbed 'coronal lines' in view of the presence of similar emission in the spectrum of the solar corona [1]. Spectral lines identified in Seyferts include [Fe VII] $\lambda\lambda$ 5721, 6087 Å, [Fe X] $\lambda$ 6374 Å, [Fe XI] $\lambda$ 7892 Å and [Fe XIV] $\lambda$ 5303 Å [2]. Studies of the highly ionised iron spectral line profiles have revealed subtle differences between these features and the prominent narrow lines associated with [O I], [O II], [O III], [N II] and [S II] [3]. In particular, the iron lines have been found to have a greater width and a Doppler shift compared to the remainder of the narrow line spectrum, and the magnitude of this discrepancy correlates with the degree of ionisation [4]. This has traditionally been interpreted as evidence of stratification of the narrow line region, with the iron emitting zone, particularly where the ionisation is greatest, presumed to be the innermost layer most exposed to photoionisation by high energy radiation generated in the vicinity of the AGN accretion disk surrounding a central black hole [5,6,7].

The Sloan Digital Sky Survey (SDSS) [8] has greatly increased the number of known AGN, and has unearthed a variety of unusual objects. One of these is SDSS J0952+2143, which, in addition to a peculiar multiple Balmer line profile, exhibited coronal lines with relative strengths a multiple of what has been hitherto observed [9,10]. The SDSS spectrum of this object recorded in December 2005 displays a [Fe VII] $\lambda$ 6087 Å line strength of as much as 67% of the [O III] $\lambda$ 5007 Å flux (even amongst AGN recognised as having strong coronal emission, this ratio is typically only 4% [7]). Follow-up spectroscopic observations in 2008 showed that the relative strength of the coronal lines in

SDSS J0952+2143 had by then decreased by about 50% for [Fe VII] and 80%-90% for higher levels of ionisation [10]. This unusual coronal line brightening event illustrates that processes other than photoionisation of narrow line region gas from a luminous central source may be responsible for the coronal emission, including exceptional supernova eruptions or tidal disruptions due to galaxy interaction [10]. This raises the question whether other data exists displaying similar hyper-strong coronal line emission.

This papers presents SDSS J1055+5637, another AGN with an SDSS spectrum showing coronal lines way stronger than normal [11], though not quite on the scale seen for SDSS J0952+2143.

## 2. Spectral analysis

This paper investigates archival wavelength and flux calibrated spectra acquired online from the SDSS database [8]. Figure 1 displays the SDSS spectrum of SDSS J1055+5637, where the abnormally strong coronal lines are immediately evident.

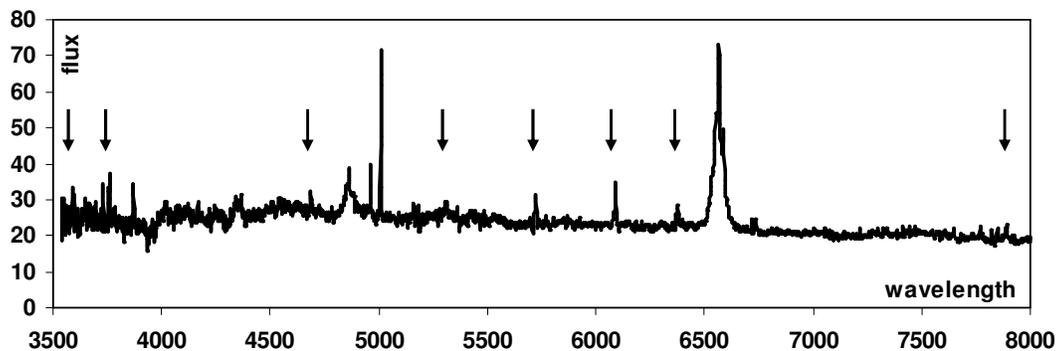

**Figure 1.** SDSS spectrum of SDSS J1055+5637. The flux is given in units of $\times 10^{-17}$ erg s$^{-1}$ m$^{-2}$ Å$^{-1}$ and the wavelength in Å. The arrows pinpoint the locations of the unusually strong coronal lines.

Continuum levels $C(\lambda)$ in the spectral neighbourhood of an emission line were estimated through the fitting of a power law of the form $C(\lambda) = C_0 \lambda^{-\alpha}$ between two continuum points identified 'by eye' on either side of the line. The line flux $F$ was hence determined to be the difference between the total and continuum fluxes in the wavelength range flanked by these points. A Gaussian or, where line blending was evident, several Gaussians were then fitted to the line profile (after subtraction of the continuum). Balmer broad line components were modelled by Lorentzian profiles rather than Gaussians. Figure 2 illustrates the outcome of the fitting of the spectral region around the hydrogen alpha line ($\lambda$ 6563 Å). This fit was also performed 'by eye', constraining the three narrow lines to all be of the same width (in terms of Doppler velocities) and fixing the ratio between the two [N II] lines to be 3:1, which is a close approximation of the theoretical value [12]. This approach has been validated by the good match between the measured and modelled profiles.

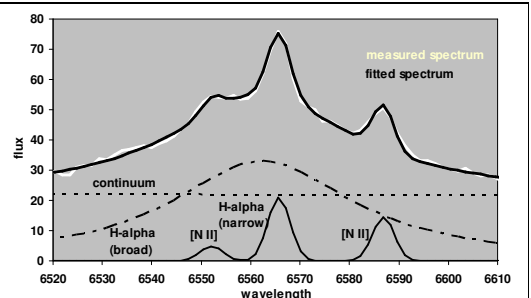

**Figure 2.** Section of the spectrum in the vicinity of Hα for SDSS J1055+5637. The white line represents the measured spectrum, while the thick solid black line represents the fitted model. The components of this model are also shown in the figure as broken lines for the continuum and broad component and thin solid lines for the three narrow lines.

Emission line widths were always determined to be greater than the instrumental resolution of ~150 km s$^{-1}$ characteristic of SDSS spectra. In particular, the coronal line *FWHM* (full width at half maximum) typically exceeded the corresponding value for emission lines such as [O III] by about 100 km s$^{-1}$. The coronal lines are furthermore blue-shifted by on average 75 km s$^{-1}$ relative to [O III].

The flux ratios between specific emission lines are indicative of the physical conditions in the gas they originate in. In particular, specific [O III] and [S II] line ratios enable the estimation of the gas density and temperature [13]. Where photoionisation is the dominant emission mechanism, the ratios between iron emission lines from different degrees of ionisation provide clues about the system geometry and physical processes [5]. Finally, this work also includes a determination of the asymmetry parameter *A20* [14] for the [O III].

In addition to SDSS J1055+5637, for comparison purposes, this study also examines the spectra of three AGN with distinct coronal line emission, and one Seyfert with no iron emission lines detectable. Furthermore, the line ratios and some line parameters of SDSS 0952+2143 have been redetermined to contrast it with the primary object of this study. Table 1 lists the coordinates, redshift, reddening *E*(*B*–*V*) at those coordinated by the interstellar medium (ISM) in our Galaxy (obtained from Schlegel et al [15]) and the date the original spectrum was recorded. Note that the redshifts of all six objects in Table 1 are similar, meaning that the size of the region captured within a slit is similar in all cases. The spectra of two of these additional objects are displayed in Figure 3 (the remaining ones may be inspected on the SDSS website or in the online ZORROASTER catalogue [16]).

**Table 1**. Details of Seyfert galaxies described in this paper.

| Object | R.A. (2000) | Dec. (2000) | *z* (SDSS) | *E*(*B*–*V*) (ISM) | Date of SDSS observation |
|---|---|---|---|---|---|
| SDSS J1055+5637 | 10h55m26.4s | +56°37′13″ | 0.0748 | 0.008 | 9 Apr 2002 |
| SDSS J0952+2143 | 09h52m09.6s | +21°43′13″ | 0.0790 | 0.028 | 30 Dec 2005 |
| SDSS J1355+6440 | 13h55m42.8s | +64°40′45″ | 0.0754 | 0.013 | 16 Mar 2001 |
| SDSS J1429+4518 | 14h29m25.1s | +45°18′32″ | 0.0749 | 0.012 | 30 Mar 2003 |
| SDSS J2356–1016 | 23h56m54.3s | –10°16′05″ | 0.0741 | 0.032 | 22 Aug 2001 |
| SDSS J1314+2606 | 13h14m47.1s | +26°06′24″ | 0.0720 | 0.015 | 28 Feb 2006 |

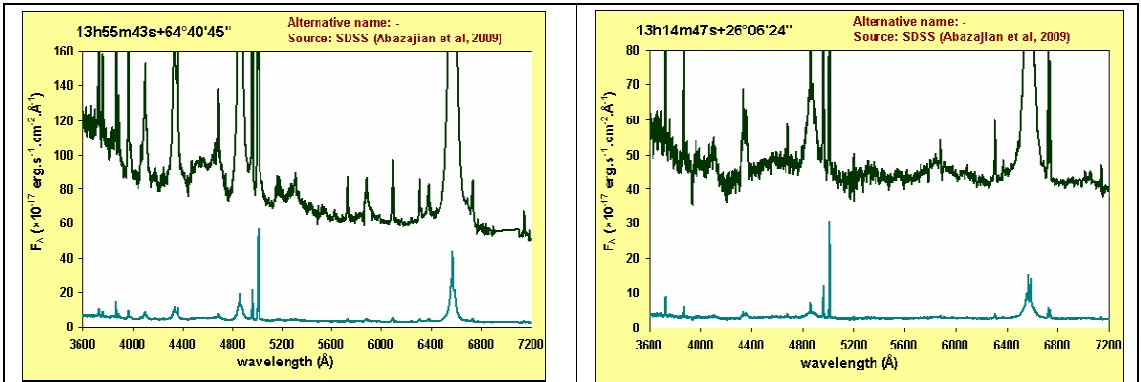

**Figure 3(a).** Spectrum of SDSS J1355+6440. The vertical scale refers to the top spectrum. The lower spectrum is a scaled down version of the top one.

**Figure 3(b).** Spectrum of SDSS J1314+2606. As in Figure 2(a), the lower spectrum is a scaled down version of the upper one.

Table 2 lists the line fluxes of all emission features of the entire AGN sample studied here. Based on the line strength uncertainties available at SDSS for each spectrum evaluated here, flux errors will be of the order of 5% for strong lines down to about 20% for the weaker emission features. Any corrections due to underlying absorption will only affect higher order Balmer lines, and will even then only be minor in view of the virtual invisibility of most features associated with the galaxy background component (e.g. Na I).

Table 3 then presents the results of the determination of a variety of parameters obtained from the line profile fitting and the analysis of the line flux ratios. The full width at half maximum of the hydrogen beta line (*FWHM-β*) indicates the broad line region gas velocities (projected towards the observer). The narrow line widths (*FWHM-n*) and asymmetry parameters quoted in the table are those of [O III] $\lambda$ 5007 Å, which is strong in all spectra. The reddening parameter $E(B–V)$ was determined from the flux ratio of Hα to Hβ. This was done separately from the broad and narrow components. Column 7 records the [Fe X] $\lambda$ 6374 Å to [Fe VII] $\lambda$ 6087 Å line flux ratio, which best illustrates how different ionisation levels are represented in each of our AGN sample. The last columns list the free electron density $n_e$ deduced from the [S II] $\lambda\lambda$ 6716,6731 Å doublet and the free electron temperature $T_e$ calculated from the relative strength of [O III] $\lambda$ 4363 Å compared to the other [O III] lines.

**Table 2**. Line fluxes of the objects investigated in this study (in units of $10^{-15}$ erg s$^{-1}$ m$^{-2}$). An (↑) indicates the presence of the line, but that it could not be measured with accuracy due to it being blended with the line above.

| $\lambda$ (Å) | Identification | SDSS 1055+5637 | SDSS 0952+2606 | SDSS 1355+6440 | SDSS 1429+4518 | SDSS 2356–1016 | SDSS 1314+2606 |
|---|---|---|---|---|---|---|---|
| 3587.2 | [Fe VII] | 0.50 | 0.61 | 1.60 | - | 4.15 | - |
| 3727 | [O II] | 0.57 | 0.81 | 6.18 | 3.48 | 18.2 | 5.38 |
| 3759.9 | [Fe VII] | 0.92 | 1.31 | 3.40 | 0.58 | 6.29 | - |
| 3868.74 | [Ne III] | 0.64 | 0.56 | 7.86 | 2.28 | 9.33 | 2.41 |
| 4101.73 | Hδ | - | - | 19.0 | 7.05 | 10.3 | 3.32 |
| 4340.46 | Hγ | 2.14 | 0.87 | 46.8 | 12.2 | 22.8 | 7.87 |
| 4363.21 | [O III] | ↑ | ↑ | 4.18 | - | 4.04 | 0.72 |
| 4550 | Fe II band | weak | - | 50.6 | 14.8 | 22.6 | - |
| 4685.68 | He II | 0.50 | 1.18 | 4.06 | 0.94 | 4.96 | 0.76 |
| 4861.33 | Hβ broad | 5.08 | 2.02 | 92.7 | 27.0 | 66.2 | 20.8 |
| 4861.33 | Hβ narrow | 0.30 | 0.43 | 5.54 | 1.78 | 7.68 | 1.57 |
| 4958.91 | [O III] | 0.85 | 0.52 | 26.2 | 8.07 | 31.5 | 8.85 |
| 5006.84 | [O III] | 2.40 | 1.61 | 76.5 | 25.6 | 89.4 | 26.1 |
| 5250 | Fe II band | weak | - | 26.0 | - | 13.2 | - |
| 5302.86 | [Fe XIV] | 0.75 | 1.05 | 1.00 | - | - | - |
| 5721.11 | [Fe VII] | 0.73 | 1.04 | 2.27 | - | 3.79 | - |
| 5875.62 | He I | - | - | 7.23 | 5.74 | 7.38 | 0.92 |
| 6086.92 | [Fe VII] | 1.23 | 1.44 | 3.79 | 1.24 | 6.27 | - |
| 6300.30 | [O I] | - | - | 2.61 | 1.26 | 2.83 | 1.38 |
| 6374.51 | [Fe X] | 0.75 | 2.46 | 2.83 | 0.47 | 4.87 | - |
| 6562.80 | Hα broad | 19.5 | 16.5 | 302 | 126 | 243 | 82.4 |
| 6562.80 | Hα narrow | 1.26 | 1.75 | ↑ | 10.1 | 24.0 | 7.51 |
| 6583.45 | [N II] | ↑ | ↑ | ↑ | 10.6 | 22.2 | 8.56 |
| 6716.44 | [S II] | 0.22 | 0.49 | 1.58 | 2.32 | 7.43 | 3.02 |
| 6730.82 | [S II] | 0.25 | 0.43 | 2.03 | 1.79 | 6.42 | 2.66 |
| 7135.8 | [Ar III] | - | - | 1.39 | 0.59 | 2.17 | 0.65 |
| 7325 | [O II] | - | - | 2.39 | - | 2.10 | - |
| 7891.94 | [Fe XI] | 0.72 | 1.76 | no data | - | 3.13 | - |

**Table 3.** Parameters determined from the spectral properties.

| Object | FWHM-$\beta$ [km s$^{-1}$] | FWHM-n [km s$^{-1}$] | A20 [O III] | E(B–V) (broad) | E(B–V) (narrow) | F(6374)/ F(6087) | $n_e$ (cm$^{-3}$) | $T_e$ (K) |
|---|---|---|---|---|---|---|---|---|
| SDSS J1055+5637 | 2170 | 229 | 0.060 | 0.21 | 0.30 | 0.61 | 700 | - |
| SDSS J0952+2143 | 1700 | 351 | 0.173 | 0.95 | 0.25 | 1.70 | 300 | - |
| SDSS J1355+6440 | 2350 | 284 | 0.241 | 0.38 | - | 0.76 | 1000 | 30000 |
| SDSS J1429+4518 | 1850 | 172 | 0.095 | 0.60 | 0.40 | 0.38 | 100 | - |
| SDSS J2356–1016 | 2350 | 337 | 0.148 | 0.14 | 0.00 | 0.77 | 300 | 25000 |
| SDSS J1314+2606 | 2960 | 256 | 0.147 | 0.23 | 0.43 | - | 300 | 18000 |

## 3. Discussion

### 3.1. SDSS J1055+5637

The measurements in this work confirm that the coronal emission lines in SDSS J1055+5637 are indeed exceptionally strong (e.g. $F(6087)/F(5007) = 0.51$). The ratios of coronal line fluxes to those of the strongest [O III], [O II] and [O I] emission features in this object exceed those determined in other AGN with coronal lines [7, 11] (with the exception of SDSS J0952+2143).

If one does not consider the coronal lines, SDSS J1055+5637 has a very typical Seyfert 1.5 spectrum. The widths, line profiles and gas characteristics are normal for the class. The broad line component is blue-shifted by 150 km s$^{-1}$ compared to the corresponding narrow lines, highlighting a systemic motion of the broad line region gas.

The most plausible explanation for the unusual strength of the coronal lines is that this effect is due to an earlier very bright phase of the inner nuclear zone consisting of the accretion disk and its immediate surroundings. The possible sequence of events thereafter is represented in Figure 4. The light echo from this outburst would pass through the broad line region within a few days, causing this zone to brighten (a). Thereafter the light burst reaches the innermost parts of the narrow line region, where the coronal lines are believed to originate, and which would then result in a brightening of that zone as seen in the spectrum investigated (b). Some time after that the emission from that zone again fades as the pulse reaches further parts of the narrow line region (c).

It is probable that spectra of this object from a different epoch would, if they were available, display rather different coronal line emission features. Future spectroscopy of SDSS J1055+5637 is therefore strongly encouraged.

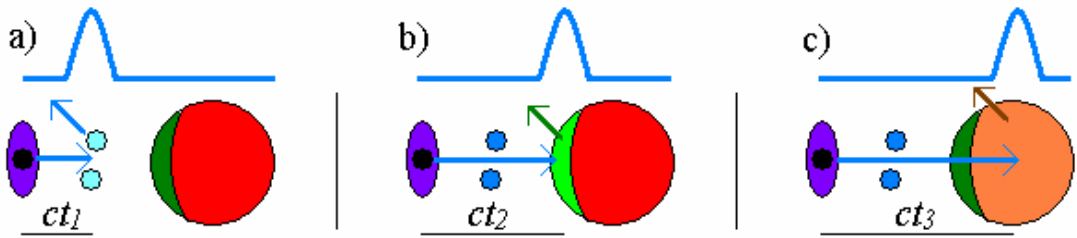

**Figure 4.** Illustration of the reverberation of a light pulse generated in the accretion disk (at the left of each of the diagrams). The diagrams are not drawn to scale. The pulse reaches the fast-moving broad line gas clouds in (a), the coronal line emitting zone at the inner part of the narrow line region in (b), and the other parts of the narrow line region in (c).

### 3.2. Other objects

SDSS J0952+2143 has been comprehensively studied in the past [9,10], and will not be discussed in depth here. Note that the high $F(6374)/F(6087)$ ratio was only a temporary phenomenon presumably associated with the immediate aftermath of an eruptive event, and that for subsequent spectra this ratio was much closer to the other values in Table 3 [10].

SDSS J1355+6440 (also referred to as VII Zw 533), SDSS J1429+4518 and SDSS J2356–1016 are more conventional broad-line Seyferts with coronal line spectra of varying prominence. The first mentioned of these displays a relatively large narrow line asymmetry and a greater electron density.

SDSS J1314+2606 (also known as Arp 60) is characterised by relatively wide broad lines, with the width of the He I line in Figure 3(b) apparently exceeding that of the Balmer lines.

## 4. Conclusion

SDSS J1055+5637 has been confirmed to be a Seyfert galaxy with the second-strongest set of coronal emission lines identified to date. It is postulated that these unusually strong features are due to a light echo transiting the coronal line formation zone following a powerful brightening of the inner nucleus. A later epoch spectrum would be required to confirm this hypothesis.

**Acknowledgments –** This paper utilized data from the Sloan Digital Sky Survey (SDSS). Funding for the SDSS and SDSS-II has been provided by the Alfred P. Sloan Foundation, the Participating Institutions, the National Science Foundation, the U.S. Department of Energy, the National Aeronautics and Space Administration, the Japanese Monbukagakusho, the Max Planck Society, and the Higher Education Funding Council for England. The SDSS Web Site is http://www.sdss.org/.

The SDSS is managed by the Astrophysical Research Consortium for the Participating Institutions. The Participating Institutions are the American Museum of Natural History, Astrophysical Institute Potsdam, University of Basel, University of Cambridge, Case Western Reserve University, University of Chicago, Drexel University, Fermilab, the Institute for Advanced Study, the Japan Participation Group, Johns Hopkins University, the Joint Institute for Nuclear Astrophysics, the Kavli Institute for Particle Astrophysics and Cosmology, the Korean Scientist Group, the Chinese Academy of Sciences (LAMOST), Los Alamos National Laboratory, the Max-Planck-Institute for Astronomy (MPIA), the Max-Planck-Institute for Astrophysics (MPA), New Mexico State University, Ohio State University, University of Pittsburgh, University of Portsmouth, Princeton University, the United States Naval Observatory, and the University of Washington.

## References
[1] Oke J B, Sargent W L W 1968 *Astrophys. J.* **151** 807
[2] Penston M V, Fosbury R A E, Boksenberg A, et al. 1984 *Mon. Notices Roy. Astron. Soc.* **208** 347
[3] Appenzeller I, Östreicher R 1988 *Astron. J.* **95**, 45
[4] Grandi S 1978 *Astrophys. J.* **221** 501
[5] Korista K T, Ferland G J 1989 *Astrophys. J.* **343** 678
[6] Rodriguez-Ardila A, Prieto M A, Viegas S, Gruenwald R 2006 *Astrophys. J.* **653** 1098
[7] Gelbord J M, Mullaney J R, Ward M J 2009 *Mon. Notices Roy. Astron. Soc.* **397** 172
[8] Ahn C P, Alexandroff R, Allende Prieto C, et al. 2014 *Astrophys. J. Suppl.* **211** A17. SDSS data accessible at http://skyserver.sdss.org/
[9] Komossa S, Zhou H, Wang T, et al. 2008 *Astrophys. J.* **678** L13
[10] Komossa S, Zhou H, Rau A, et al. 2009 *Astrophys. J.* **701** 105
[11] Nagao T, Taniguchi Y, Murayama T 2000 *Astron. J.* **119** 2605
[12] Storey P J, Zeippen C J 2000 *Mon. Notices Roy. Astron. Soc.* **312** 813
[13] Osterbrock D E, Ferland G J 2005, *Astrophysics of Gaseous Nebulae and Active Galactic Nuclei*, 2nd Ed., ISBN 978-1-891389-34-4, University Science Books
[14] Whittle M 1985 *Mon. Notices Roy. Astron. Soc.* **213** 1
[15] Schlegel D J, Finkelbeiner D P, Davis M 1998 *Astrophys. J.* **500** 525
[16] Winkler H 2013 *ZORROASTER online catalogue*, http://www.uj.ac.za/EN/Faculties/science/departments/physics/Pages/ZORROASTER.aspx